\documentclass[12pt]{iopart}

\usepackage{graphicx}
\usepackage{color}
\usepackage{changes} 
\usepackage{xfrac} 
\usepackage{hyperref}
\usepackage{xcolor}
\usepackage[width=1.0\textwidth,font=scriptsize]{caption}

\begin{document}
	
\title{Novel single-layer vanadium sulphide phases}
\author{Fabian Arnold$^{1}$, Raluca-Maria Stan$^{1}$, Sanjoy K. Mahatha$^{1}$, H. E. Lund$^{1}$, Davide Curcio$^{1}$, Maciej Dendzik$^{1}$, Harsh Bana$^2$, Elisabetta Travaglia$^2$, Luca Bignardi$^3$, Paolo Lacovig$^3$, Daniel Lizzit$^3$, Zheshen Li$^{1}$, Marco Bianchi$^{1}$, Jill A. Miwa$^{1}$, Martin Bremholm$^4$, Silvano Lizzit$^3$, Philip Hofmann$^{1}$, and C. E. Sanders$^{1}$}
\address{$^1$Department of Physics and Astronomy, Interdisciplinary Nanoscience Center (iNANO), Aarhus University, 8000 Aarhus C, Denmark.}
\address{$^2$Department of Physics, University of Trieste, Via Valerio 2, 34127 Trieste, Italy.}
\address{$^3$Elettra - Sincrotrone Trieste S.C.p.A., AREA Science Park,
Strada Statale 14, km 163.5, 34149 Trieste, Italy.}
\address{$^4$Department of Chemistry, Aarhus University, 8000 Aarhus C, Denmark.}
\eads{\mailto{sanders.charlotte@phys.au.dk}}
\date{\today}
	
	
\begin{abstract}
VS$_2$ is a challenging material to prepare stoichiometrically in the bulk, and the single layer has not been successfully isolated before now.  Here we report the first realization of single-layer VS$_2$, which we have prepared epitaxially with high quality on Au(111) in the octahedral (1T) structure. We find that we can deplete the VS$_2$ lattice of S by annealing in vacuum so as to create an entirely new two-dimensional compound that has no bulk analogue.  The transition is reversible upon annealing in an H$_2$S gas atmosphere.  We report the structural properties of both the stoichiometric and S-depleted compounds on the basis of low-energy electron diffraction, X-ray photoelectron spectroscopy and diffraction, and scanning tunneling microscopy experiments.
\end{abstract}

\maketitle

\section{Introduction}

Many transition metal dichalcogenides (TMDCs) have by now been subjected to intensive investigation in both their bulk and single-layer (SL) forms.  The bulk compounds, many of which are characterized by a layered structure, exhibit a complex array of symmetry-breaking collective ground states.  Meanwhile, in the SL limit, quantum confinement and reduction of symmetry lead to significant changes in the electronic properties relative to those of the corresponding bulk parent materials \cite{Wilson:1969,Chhowalla:2013}.  For example, in the case of group-VI semiconducting TMDCs such as MoS$_2$, these changes include an indirect-direct band gap transition \cite{Chhowalla:2013,Splendiani:2010aa,Mak:2010} and coupled spin-valley physics \cite{Xiao:2012}.   Complicating the picture in the SL limit, other factors including the removal of interlayer coupling, hybridization with the substrate \cite{Bruix:2016,Sanders:2016,Wehling:2016}, and substrate screening effects \cite{Park:2009aa,Ugeda:2014aa,Antonija-Grubisic-Cabo:2015aa,Ulstrup:2016aa} can also be have a major effect on electronic properties.
 
Vanadium disulphide (VS$_2$) has received relatively little experimental investigation until now.  It is unusual among layered TMDCs in having no thermodynamically stable bulk polymorph; the bulk tends, rather, toward S-deficient forms with self-intercalated V between S layers \cite{Wilson:1969,Bensch:1993,Franzen:1975,Moutaabbid:2016}.  While the work of Ref.\cite{Moutaabbid:2016} suggests that the bulk stoichiometric phase might be realizable by high-pressure synthesis, fabrication of the bulk has so far been approached via preparation of stable alkali-intercalated compounds (NaVS$_2$, LiVS$_2$), from which near-stoichiometric VS$_2$ is obtained by deintercalation of the alkali metal \cite{Murphy:1977}.  The resulting bulk crystals assume the octahedral (1T) structure and readily desulphidize at temperatures as low as 570~K \cite{Murphy:1977}. Thus, bulk VS$_2$ already presents considerable experimental challenges.

This is still more true of the SL than of the bulk:  the difficulty of preparing stoichiometric bulk crystals implies that bulk exfoliation \cite{Joensen:1986,Novoselov:2005,Coleman:2011} is not a simple approach to attaining thin stoichiometric VS$_2$. Moreover, like other group-V metallic TMDCs \cite{Sanders:2016}, VS$_2$ is chemically reactive (\textit{e.g.,} \cite{Feng:2012}), and exfoliated thin layers would not be expected to be air-stable. The isolation of SL VS$_2$ has therefore not been achieved, even though thin layers can be exfoliated \cite{Feng:2012,Feng:2011}. However, SL VS$_2$  has been the focus of considerable theoretical interest, especially because of the possibility that there are intriguing magnetic properties in this material. In particular, SL VS$_2$ in the trigonal prismatic (``1H'') structural phase has been predicted to be a ferromagnetic semiconductor \cite{Zhuang:2016,Isaacs:2016} with strain-tunable magnetic moment and coupling \cite{Ma:2012}, and is a candidate for strongly correlated electron physics \cite{Isaacs:2016}. Meanwhile, the SL 1T structural phase, which calculations indicate to be metallic and ferromagnetic \cite{Isaacs:2016,Kan:2015,Zhang:2013}, may manifest charge density waves (CDWs):  the bulk 1T analogue exhibits a CDW transition at the relatively high temperature of $T_{CDW}$~$\approx$~305~K \cite{Tsuda:1983aa,Mulazzi:2010,Sun:2015}. With examples of magnetic two-dimensional materials being still very rare \cite{Gong:2017aa,Huang:2017}, and in light of the very recent finding of ferromagnetism in SL VSe$_2$ \cite{Bonilla:2018aa}, any ferromagnetism in a SL would be highly interesting, especially if not present in the parent bulk material. SL VS$_2$ can thus be expected to be a particularly interesting two-dimensional material, as it is likely to be a future playground for studying both magnetism and correlated behavior in the SL limit.  

In this paper, we report the synthesis of epitaxial SL VS$_2$ on Au(111) and show that the initially synthesized 1T phase can be transformed into two other structures by annealing in vacuum. This process is interpreted as caused by S loss and can, in fact, be reversed by annealing the sample in a background pressure of H$_2$S. The paper is organised as follows:  After briefly describing the methods in the next section, we describe the three observed V sulphide structures on Au(111) and the transitions between them, as observed by low-energy electron diffraction (LEED) and scanning tunneling microscopy (STM). We then report the S~2p and V~3p core level spectra for the three structures and provide a detailed structural determination using X-ray photoelectron diffraction (XPD).

\section{Methods}

Samples were synthesized according to the methods described in Refs. \cite{Sanders:2016,ProfSlacker:2015,Gronborg:2015,Dendzik:2015,Fuchtbauer:2014}. Starting from a clean Au(111) surface, as judged by the presence of the herringbone reconstruction and a narrow electronic surface state, V was evaporated onto the surface at room temperature, either from an electron-beam evaporator (V rod of 99.8{\%} purity from Goodfellow) or from a hot V wire ({99.8{\%} purity from Goodfellow). Subsequently, the sample was annealed in an atmosphere of sulphiding gas (hydrogen sulphide (H$_2$S) at  a background pressure of $\approx10^{-4}$~mbar  or dimethyldisulphide (DMDS) at a background pressure of $\approx10^{-6}$~mbar) to temperatures in the range 670--705~K (base pressure $\approx$10$^{-10}$~mbar). As an alternative approach, when H$_2$S was used as the sulphiding agent the V evaporation was in some cases carried out with the surface held already at the annealing temperature. All of these approaches lead to samples with very similar properties, as judged by STM, LEED, and X-ray photoelectron spectroscopy (XPS).  The finding that samples grown with DMDS as a S source have similar quality to those grown with H$_2$S is consistent with previous studies of MoS$_2$ growth \cite{Fuchtbauer:2014}. Samples were prepared and probed \textit{in situ} in ultra-high vacuum (UHV) conditions, because we observed a strong sensitivity of the material to air.

LEED experiments were performed at both low temperature (35~K) and room temperature, with the same results in both cases.  In-plane lattice parameters of the V sulphide structures were determined by a calibration to the known lattice constant of Au. For STM measurements, a home-built Aarhus-style STM was used at room temperature \cite{Laegsgaard:2001}, and measurements were calibrated to atomic-resolution data from Au(111) and from graphene.

XPS measurements were made at room temperature.  Comparative XPS measurements of H$_2$S- and DMDS-grown samples were performed at the MatLine of ASTRID2, while detailed XPS and XPD studies of H$_2$S-grown samples were performed at the SuperESCA beamline of ELETTRA \cite{SMAT,Baraldi:2003}. XPD patterns were measured by collecting XPS spectra for more than 1000 different polar ($\theta$) and azimuthal ($\phi$) angles. For each of these spectra, a peak fit analysis was performed  and the intensity $I(\theta$, $\phi)$ of each component resulting from the fit (\textit{i.e.} the area under the photoemission line) was extracted. The resulting XPD patterns are the stereographic projection of the modulation function $\chi$, which was obtained from the peak intensity for each emission angle  $I(\theta, \phi) $ as \[{\chi} = \frac{I(\theta, \phi) - I_0(\theta)}{I_0(\theta)}\] where $I_0$($\theta$) is the average intensity for each azimuthal scan. Structural determinations were performed by comparing measured XPD patterns to multiple-scattering simulations for trial structures. Such simulations were carried out using the program package for Electron Diffraction in Atomic Clusters (EDAC) \cite{Garcia-de-Abajo:2001aa}. The agreement between the calculated modulation function $\chi_{th}$ and the measured modulation function $\chi_{ex}$ is quantified by a reliability-factor ($R$-factor). When the $R$-factor is defined as
\begin{equation}\label{chiquadrat}
R=\frac{\sum_{i}{(\chi_{th,i}-\chi_{ex,i})^2}}{\sum_{i}{(\chi_{th,i}^2+\chi_{ex,i}^2)}}, 
\end{equation}
it can have values between 0 and 2, with 0 corresponding to perfect agreement \cite{Woodruff:2007aa}. In order to identify the three structural phases reported in this paper, simulated crystal structures and lattice parameters were systematically varied in order to minimize the $R$-factor.

\section{Results}

\subsection{Three Crystalline Phases}

STM and LEED measurements reveal three distinct crystalline phases, which are shown in Fig.~\ref{fig:STM-LEED}.    We refer to the as-grown phase as ``phase~I''; it is shown in Fig.~\ref{fig:STM-LEED}(a).  The STM topography exhibits a well-ordered hexagonal moir\'{e} superstructure (moir\'{e} superlattice parameter 22(2){\AA}) in triangular SL islands, much like what has been seen previously in similarly prepared SL WS$_2$, MoS$_2$ and TaS$_2$ on Au(111) \cite{Sanders:2016,ProfSlacker:2015,Gronborg:2015,Dendzik:2015}.  At coverage close to a complete monolayer, as in Fig.~\ref{fig:STM-LEED}(a), the islands merge together, typically without domain boundaries.  An atomic-resolution image is shown as an inset and reveals a hexagonal atomic-scale structure.  On the basis of this STM data, the atomic lattice parameter is 3.0(3){\AA}.  We conclude that the islands are only one layer in thickness, on account of both the apparent height with respect to the Au surface---although this alone is not decisive, since the apparent height depends not only on the island height but also on the density of states---and on the strong appearance of the moir\'{e} superstructure, which we find to be nearly absent in the bilayer, consistent with observations in related systems (\textit{e.g.,} \cite{Gronborg:2015}).  In the case of the islands shown in the main panel of Fig.~\ref{fig:STM-LEED}(a), the apparent height is 1.6(2)~\AA{} for the scan parameters used (see caption).  (For measurement of the actual geometric height of the layer, see Section \ref{XPDSubsection} below.)  The LEED pattern (Fig.~\ref{fig:STM-LEED}(a), bottom) reveals diffraction maxima for both the Au(111) substrate and for the V sulphide SL;  the reciprocal lattice vectors for both are shown in the figure. We observe sharp spots indicative of a high degree of ordering over the whole sample, and moir\'{e} satellite spots with strong intensity.  Note that the moir\'{e} spots, including even those of higher orders, are significantly stronger than the weak main spots of the V sulphide atomic lattice (marked with a purple arrow in the LEED panel of Fig.~\ref{fig:STM-LEED}(a)); the reasons for this are not entirely clear, but we note that LEED intensity is a complex quantity involving multiple-scattering \cite{Rous:1986}, and that it affects spot intensity in non-trivial ways.  The hexagonal atomic lattice (purple arrow) is rotationally aligned with the atomic lattice of the underlying Au(111) (yellow arrow in the panel), a fact which suggests nontrivial substrate interaction.  LEED indicates that the moir{\'e} superstructure has a lattice parameter of 25(7)~{\AA}, while the atomic lattice parameter is 3.25(10)~{\AA}, in agreement with the STM results.  Neither the STM nor LEED technique is capable of distinguishing between the 1H and 1T configurations of the SL, but the atomic lattice constant observed here is in agreement with that of the 1T polymorph of bulk crystalline VS$_{2}$ \cite{Moutaabbid:2016,Murphy:1977}. The lattice parameters do not match those of any other V sulphide compound of which we are aware in the literature; more information about such compounds is included in \cite{SMAT}.  Note that no bulk 2H structural configuration has yet been observed experimentally.  Weak streaks in the LEED pattern are marked by black arrows in Fig. \ref{fig:STM-LEED}(a) and can be presumably ascribed to periodically arranged dislocation lines.

\begin{figure*}
   \includegraphics[width=1.0\textwidth]{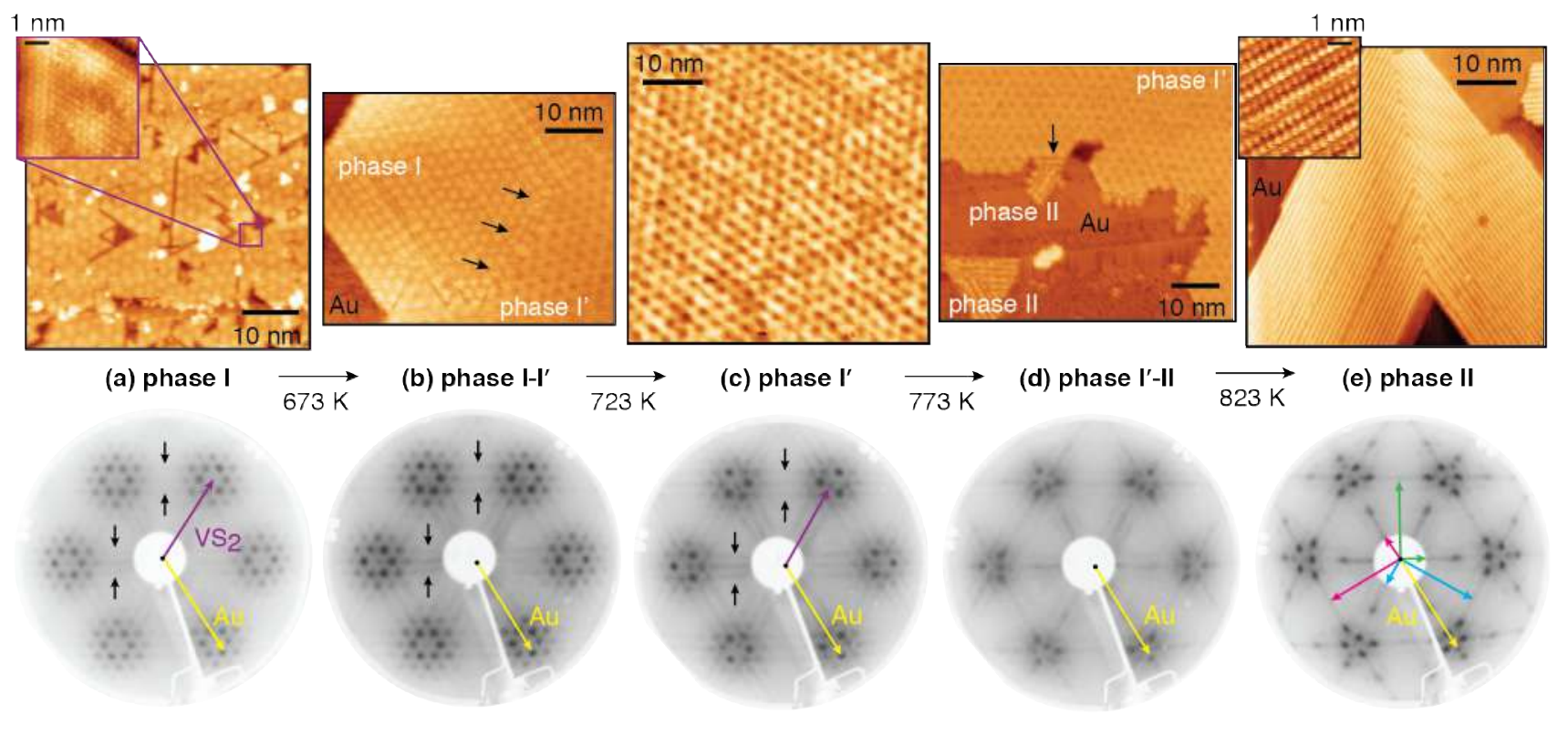}\\
   \caption{Evolution of the crystal morphology of V sulphide on Au(111) upon annealing in UHV.  The different phases and the respective annealing temperatures  are indicated. Top row shows STM data; bottom row shows corresponding LEED data.  (a) Phase~I; the purple box marks the area at which the inset atomic-resolution image was acquired.  (b) Transition between phases~I and I$'$.  (c) Intermediate phase~I$'$.  (d) Transition between phases~I$'$ and II.  (e) Phase~II; the inset atomic-resolution image and the image in the main panel were acquired from different but similar samples.  Black arrows in STM data mark grain boundaries between phases.  Black arrows in LEED data mark weak diffraction lines.  Purple arrows superimposed on LEED data indicate the reciprocal lattice vectors of phase~I in (a) and of phase I$'$ in (c); yellow arrows indicate the reciprocal lattice vector of Au(111); and green, blue, and pink arrows in (e) indicate reciprocal lattice vectors of the three domains of phase~II.   STM was acquired at room temperature with the following scanning parameters:  (a) $I_t$~=~0.30~nA, $V_B$~=~825~mV; inset of (a) $I_t$~=~1.26~nA, $V_B$~=~300~mV; (b) $I_t$~=~-0.26~nA; $V_B$~=~656~mV; (c) $I_t$~=~-0.33~nA; $V_B$~=~-656~mV; (d) $I_t$~=~-0.48~nA; $V_B$~=~565~mV; (e) $I_t$~=~0.32~nA; $V_B$~=~1117~mV; inset of (e) $I_t$~=~1.03~nA, $V_B$~=~335~mV.  All LEED data were acquired at $T$~=~35~K with $E_K$~=~92.8~eV.}
   \label{fig:STM-LEED}
\end{figure*}

Upon annealing in UHV to 673(15)~K, there is a transition to an intermediate phase that we label ``phase~I$'$'' because of its very strong similarity to phase I.  Both phase~I$'$ and a defective version of phase~I are simultaneously present in the STM and LEED data shown in Fig. \ref{fig:STM-LEED}(b). Like phase~I, phase~I$'$ exhibits a hexagonal moir\'{e} superstructure, but the presence of the two phases side-by-side in the STM image of Fig. \ref{fig:STM-LEED}(b) reveals a different appearance in the moir\'{e} patterns and a clearly visible domain boundary that is marked by black arrows.  In Fig. \ref{fig:STM-LEED}(b), phase~I$'$ begins to exhibit a more distorted moir\'{e} pattern with clearly observable dislocation lines, presumably due to the initiation of the phase change.  Furthermore, the distinct triangular islands of Fig. \ref{fig:STM-LEED}(a), top, have now transformed into large continuous islands.  The LEED pattern when both phases are present is very similar to that of phase~I, but there is enhanced intensity in the streak lines, indicative of greater disorder and one-dimensional dislocations.

Annealing to 723(15)~K leads to a conversion of an even larger surface area to phase~I$'$, as shown in 
Fig. \ref{fig:STM-LEED}(c).  The LEED pattern (bottom panel) now shows an additional hexagonal lattice (marked by a purple arrow) with a slightly smaller reciprocal lattice constant than that of phase I. The different origin of this lattice is qualitatively seen in the LEED pattern, as the new spot is not in the centre of the three intense spots stemming from the Au(111) and the moir\'{e} spots of phase~I. The length of this purple arrow corresponds to an atomic lattice constant of 3.3(1)~{\AA}---a slight increase with respect to phase~I. No distinct moir\'{e} spots due to this new periodicity are identified, but based on the lattice constant, a moir\'{e} periodicity of 23(6)~{\AA} would be expected, consistent with the 28(3)~{\AA} observed in STM.  We emphasize that the accurate structural determination of phase~I$'$ is hindered by the complicated LEED pattern, caused by the continued presence of phase I and the additional stripes. As we shall see below, phase~I$'$ can be interpreted as a S-depleted and hence defective version of phase ~I, and while both LEED and XPD point towards an increase of the lattice parameter on average, it should be kept in mind that this is a dislocation-dominated defective structure. 

Annealing in UHV to 773(15)~K leads to a transition to yet a third phase, whose signature in STM images is its striped appearance.  This striped phase is shown in Fig. \ref{fig:STM-LEED}(d) coexisting with phase~I$'$.  We label it ``phase~II,'' as it is distinctly different from the other phases.  In the upper part of the STM image (top panel), the hexagonal moir\'{e} pattern of phase~I$'$ is visible; by contrast, a triangular island with the striped structure appears in the bottom left corner of the image, and a similar striped domain near the middle of the scan area is attached to the phase~I$'$ island with a domain boundary that is marked by a black arrow.  The simultaneous presence of two phases results in a LEED pattern that is less sharply defined, with weaker spots and more diffuse background than the other LEED patterns shown in the figure. While the three intense moir{\'e} spots are still visible, they appear weaker in intensity, and the lower-intensity moir{\'e} spots are difficult to identify against the background. Higher-order moir{\'e} spots are not clearly observable, probably as a result of disorder within the structure.

The complete transition to phase~II is seen after annealing to 823(15)~K, and is shown in Fig. \ref{fig:STM-LEED}(e).  The inset at the top shows an atomically resolved STM image revealing a rectangular unit cell whose lattice parameters are of 8.2(8)~\AA{} and 3.1(3)~\AA{}, a structure not consistent with that of any V sulphide bulk compound of which we are aware \cite{SMAT}. The unit cell contains two rows of atoms with apparent heights that differ by 0.15(2)~{\AA}. Phase~II occurs in three rotational domains oriented at 120$^{\circ}$ angles relative to one another, corresponding to the three-fold symmetry of the underlying Au(111) substrate.   The phase~II LEED data in Fig. \ref{fig:STM-LEED}(e), lower panel, exhibits a complicated diffraction pattern which at first resembles a hexagonal reciprocal lattice with Christmas-tree-like satellite spots.  However, taking into account the rectangular unit cell of the structure identified by STM, the LEED pattern can be interpreted as seen in Fig. \ref{fig:PhaseIIILEED}.  The diffraction spots deriving from the three rotational domains and from Au(111) are distinguished in Fig. \ref{fig:PhaseIIILEED}(a) by colour coding with green, blue, pink and yellow.  In Fig. \ref{fig:PhaseIIILEED}(b) a calculated grid of reciprocal unit cells for each domain obtained from the STM data is overlaid on the LEED image.  In Fig. \ref{fig:PhaseIIILEED}(c), the calculated grid from Fig. \ref{fig:PhaseIIILEED}(b) is removed, but the coloured circles identify the origin of each spot in the pattern.  The diffraction spots can thus be assigned to the corresponding rotational domains.  The lattice constants for phase~II can be determined from this to be 9.1(1)~{\AA} and 3.3(1)~{\AA}, which are in agreement with the STM measurements.

\begin{figure}
\centering
   \includegraphics[width=0.8\textwidth]{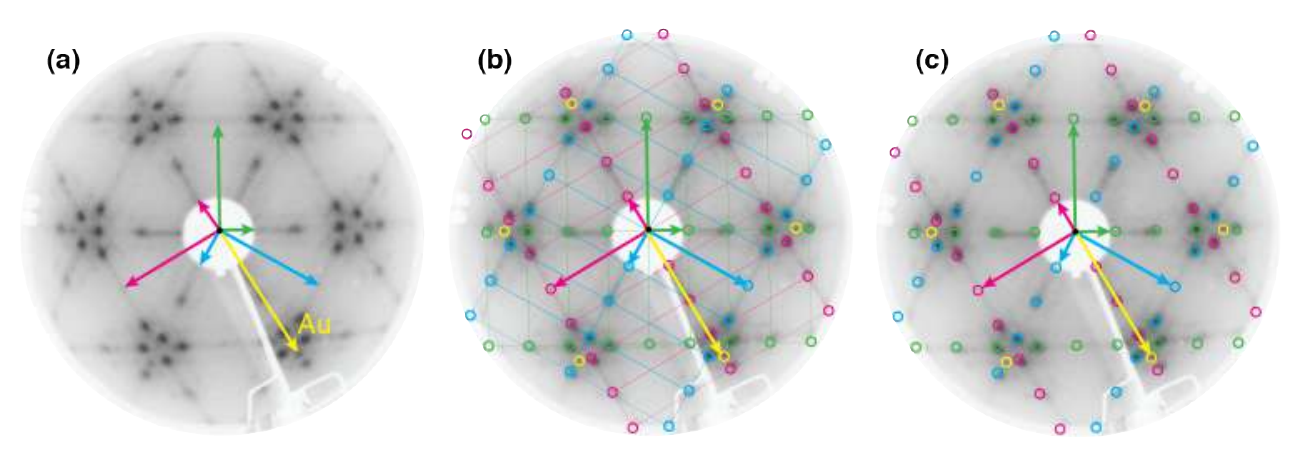}\\
   \caption{Analysis of the  LEED pattern of phase~II ($T$~=~35~K, $E_K$~=~92.8~eV).  (a) Green, blue and pink arrows indicate the reciprocal lattice vectors of the three different rotational domains, and the yellow arrow the reciprocal lattice vector of Au(111). (b) The rectangles are the reciprocal unit cells of the three rotational domains, with the colours corresponding to those in panel (a).  The circles show the positions of expected LEED spots for each domain and for the Au(111) substrate.  (c) Identical to panel (b), but with the connecting lines of the rectangular reciprocal unit cells removed, to show more clearly the agreement between the expected and observed spot positions.} 
   \label{fig:PhaseIIILEED}
\end{figure}

\subsubsection{Reversibility by Annealing in H$_2$S}

The lack of stability of bulk VS$_2$ against S loss would suggest that the observed phase changes to the SL upon annealing in UHV are likewise due to an increased depletion of S in the structures (\textit{i.e.,} a transition from VS$_2$ to V$_{1+x}$S$_{2}$). This is strongly supported by the observation that the annealing-induced phase changes can be reversed by annealing in an H$_2$S atmosphere. 

Fig. \ref{fig:PhaseReversibility} illustrates the cycle of structural phase transformations by depleting and re-supplying S. Fig. \ref{fig:PhaseReversibility}(a) shows the characteristic phase~II LEED pattern, obtained from a structure created by annealing the initially grown phase~I, as described above. This phase~II transforms into a well-defined phase I$'$ diffraction pattern when the sample is annealed to 523(15)~K in a H$_2$S background pressure of $8.5\times 10^{-5}$~mbar (Fig. \ref{fig:PhaseReversibility}(b)). For a further transformation to phase~I, both the H$_2$S background pressure and the temperature have to be increased, suggesting a more difficult integration of the S atoms into the V$_{1+x}$S$_{2}$ single layer (Fig. \ref{fig:PhaseReversibility}(b) and (c)). The resulting LEED image in Fig. \ref{fig:PhaseReversibility}(c) exhibits the typical phase~I moir\'{e} pattern and this structure can, of course, again be transformed into phases I$'$ and II by annealing in vacuum. The full reversibility of the phase changes by annealing in H$_2$S underlines the importance of S in each phase transition and supports the hypothesis that phases~I$'$ and II are S-deficient compared to phase~I. 
 
\begin{figure}
\centering
   \includegraphics[width=0.6\textwidth]{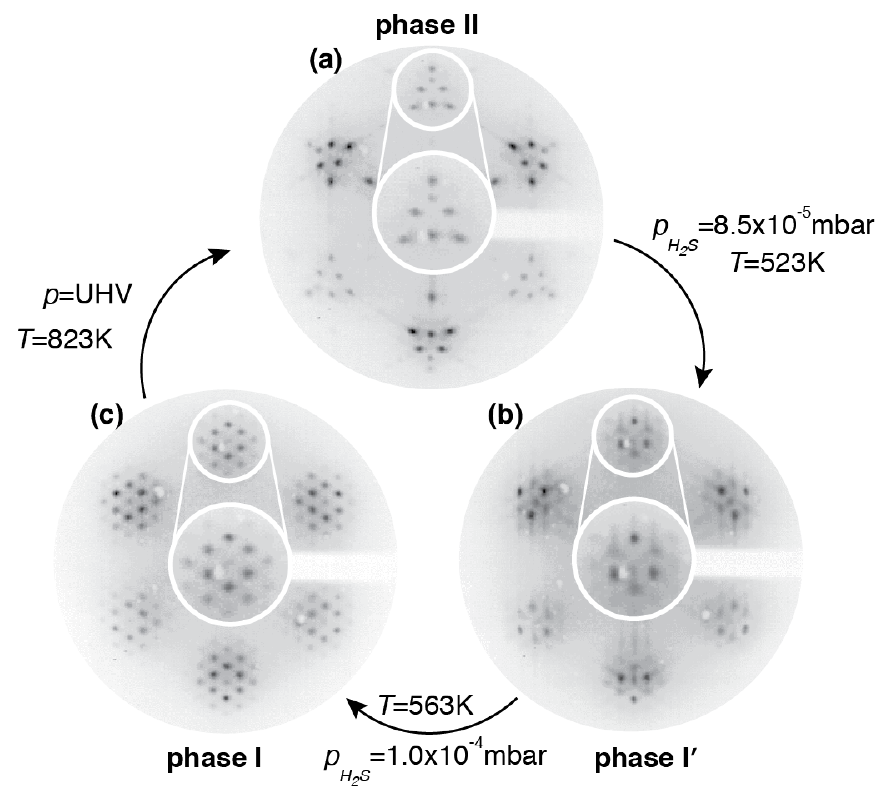}\\
   \caption{Reversibility of the three phases of V sulphide shown by the evolution of the LEED pattern ($T$~=~{room temperature}, $E_K$~=~93~eV). The different phases are indicated in the panels. The insets at the center of each panel show the magnified area around a first order spot.} 
   \label{fig:PhaseReversibility}
\end{figure}

\subsection{XPS and Structural Determination by XPD}

Additional information about the three different structures and the transformation processes can be obtained by studying high-resolution core-level spectra. XPS spectra from the S core levels would be expected to show at least two chemically different components, one from the ``top'' S atoms (at the vacuum interface) and one from the ``bottom'' S atoms (at the Au(111) interface).  Meanwhile, V atoms would not necessarily be expected---at least in the case of the simple 1T structure---to occur in more than one distinct chemical environment. In the following, we present S and V XPS data for all three structures, along with fits of the different components. Additional information about lineshapes and a table of peak positions is given in Ref. \cite{SMAT}.

From peak fitting alone it is possible neither to identify the origin of the shifted components nor to identify the detailed geometrical structure of the layer. However, such information can be obtained by XPD \cite{Woodruff:2007aa,Lizzit:1998aa}. This technique is based on emission-angle-dependent modulations of the core-level photoemission intensity from the different atoms in the layer. The intensity modulations arise from the length difference between individual scattering pathways from the emitting atom to the detector and the coherent interference of the scattered waves. The XPD modulation functions thus directly reflect the local structural environment of the emitting atom. As we shall see, the modulation functions for the top and bottom S atoms of all three structures are distinctly different from one another, and a straightforward comparison between measured and simulated modulation functions permits an unambiguous assignment of the observed peaks---for instance, to top and bottom S atoms. We will present this assignment already in connection with the discussion of the XPS peak fitting analysis, but we emphasize that it is based on the XPD results that will be presented in the subsequent section. 

\subsubsection{Peak Positions and Lineshape in XPS}

\begin{figure*}
   \includegraphics[width=1.0\textwidth]{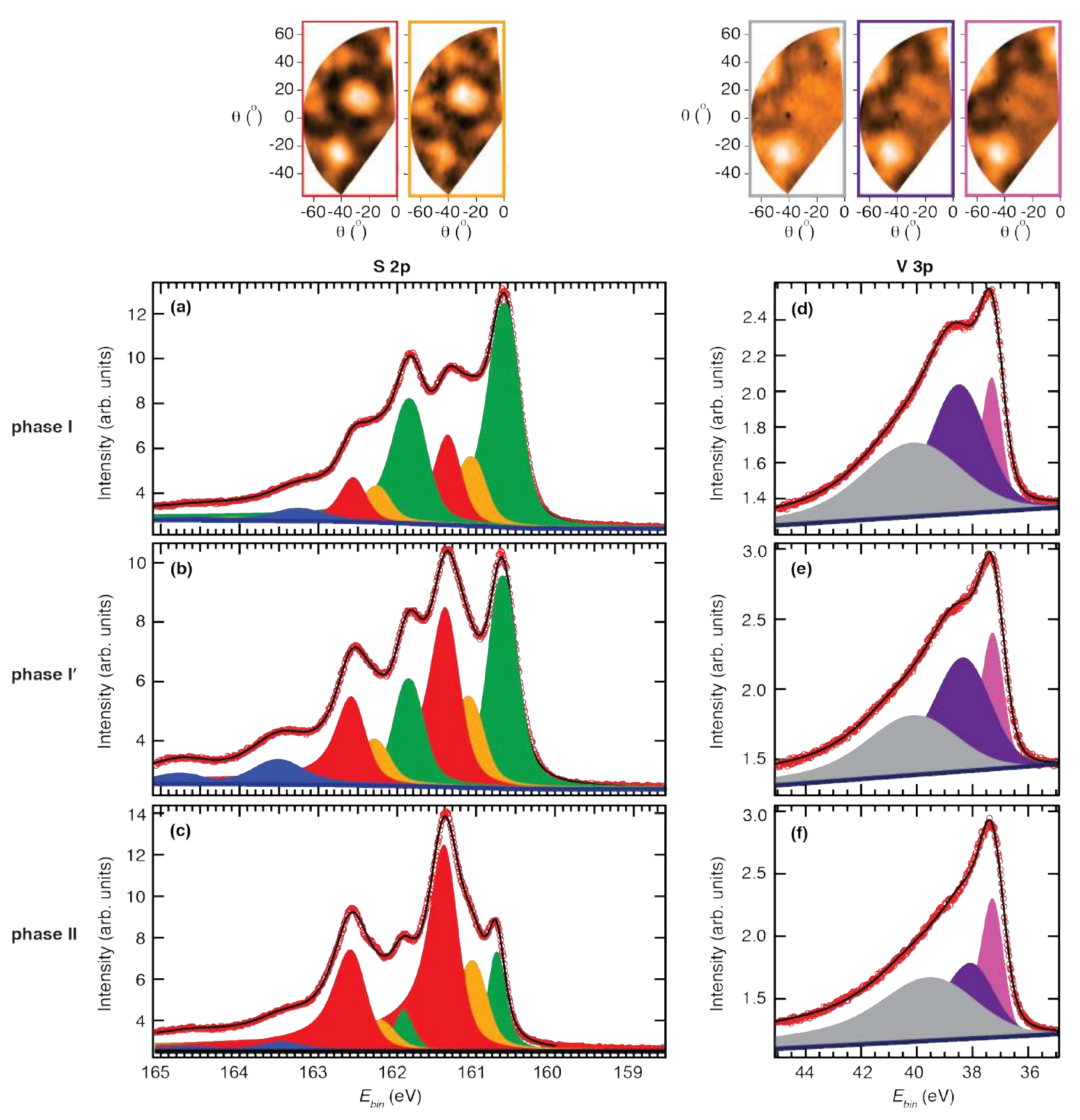}
   \caption{XPS spectra of the three phases of  V$_{1+x}$S$_{2}$ at normal emission and room temperature.  Red circles represent data points, the solid black line the fit, and the solid dark blue line the background. (a), (b), (c):  XPS spectra  for the spin-split S~2p$_{1/2}$ and 2p$_{3/2}$ core-level states ($h\nu$~=~260~eV).  Peaks are colour-coded to indicate the emitter from which each core-level state derives (see text for discussion):  green peaks originate from S atoms on top of the SL; orange and red from S atoms on the bottom of the SL.  The blue peaks are unordered sulphide species.  For same-coloured peaks, the 2p$_{3/2}$ spin-split component is at lower binding energy, the 2p$_{1/2}$ component at higher binding energy.  (d), (e), (f):  XPS spectra for the  V~3p state ($h\nu$~=~170~eV).  Top panels:  Stereographic projections of the XPS intensity modulation for phase I. The coloured frame of each panel identifies the corresponding peak component in (a) and (d). Modulation data acquired at photon energies of 400~eV for S 2p and 270~eV for V 3p.}		
   \label{fig:XPS}
\end{figure*}

Fig. \ref{fig:XPS} shows  S~2p and  V~3p core-level spectra from each of the three phases. The  S~2p spectra show clear differences for the three structural phases, and we start by analyzing their lineshape. The S~2p core level is split  due to the spin-orbit interaction into a $j=1/2$ component and a $j=3/2$ component, with a splitting of 1.20(4)~eV for all the components considered.  The spectrum for phase~I in Fig. \ref{fig:XPS}(a) requires three spin-orbit doublets to obtain a satisfactory fit:  the 2p$_{3/2}$ component of the spin-split doublet centred at a binding energy of 160.65(2)~eV (green) corresponds to the top S atoms, while those  at 161.07(2)~eV (orange) and 161.36(2)~eV (red) both derive from bottom S atoms. The assignment of both the orange and red components to S atoms in essentially the same geometric environment is confirmed by the similarity of the components' individual XPD modulation functions, which are shown in the inset at the top of the figure.  It is not entirely clear why there are two separate sets of peaks  that both derive from the bottom S layer, but a possible cause could be a variation of the S-substrate interaction within the moir\'{e} unit cell, similar to what has been observed in graphene on transition metal surfaces \cite{Presel:2015}. For graphene on Ru(0001), for instance, the C~1s peak contains multiple components from the many carbon atoms in the large moir\'{e} unit cell, but the binding energies cluster in two ranges:  this results in a spectrum that is well described by two components \cite{Alfe2013}.

To obtain a good fit in the present case, it is also required to include a lower-intensity doublet at 163.52(5)~eV (indicated with blue in the figure).  This doublet does not show any modulation in the XPD pattern.  We interpret it as being due to the presence of miscellaneous unordered sulphide species.  The peaks are probably composed of multiple weak contributions, though we are able to fit them here with a single pair of peaks.  We do not investigate them further in this study.

Upon transformation to the intermediate phase~I$'$ (Fig. \ref{fig:XPS}(b)), no significant peak shift occurs (see table in Ref. \cite{SMAT}).  The stability of the peak positions suggests that only minor changes occur in the crystal structure.  It is temping to assign the relative intensity loss of the top layer component compared to the two bottom layer components to a S depletion from the top layer; however, such a simple analysis is not unproblematic, because photoemission intensity is influenced not only by stoichiometry but also by diffraction effects that modulate core-level intensity as a result of changes in the crystal structure. The interpretation of an intensity decrease due to S loss is, however, consistent with the other observations reported in this paper and might, in view of the very minor structural changes that occur, be permissible here. 

In Fig. \ref{fig:XPS}(c), corresponding to phase II, the intensity of the top S component (green) is significantly reduced, and there is a $\approx$90~meV  shift to higher binding energy, indicating a significant change in the crystal structure---consistent with the observations from LEED and STM.  The bottom S components (red and orange) show no significant shift; however, since these peaks must consist of multiple components resulting from  the various adsorption sites on the substrate lattice, it is not simple to draw conclusions on the basis of their positions about structural modifications in the material. The blue component is reduced in intensity, as might be expected to occur simply as a result of the desorbing of disordered sulphides at elevated temperatures. 

Fig. \ref{fig:XPS}(d)-(f) shows the V~3p spectra for the three different structures.  The spectrum is very broad (the total range shown covers more than 10~eV) and requires at least three components to obtain a satisfactory fit. The unusual slope of the background is due to an Auger peak at a somewhat lower binding energy.  The spectrum for phase~I (Fig. \ref{fig:XPS}(d)) can be fit with three Doniach-{\v S}unji{\'c} functions centred at binding energies 37.32(3) (pink),  38.48(4) (purple), and 40.10(6)~eV (grey) and a linear background (dark blue line).  Na{\"i}vely, the broad, multipeak structure seems surprising, since one might expect a single component (particularly for the phase~I structure, in which---as will be shown below---there is only one V atom per unit cell).  The existence of three distinct  V~3p components seems at first to suggest the presence of V in different environments; however, this would be hard to reconcile with the very different linewidths. Moreover, all three peaks show the same XPD modulation, as seen in the top inset of the figure, and this suggests that they have an identical geometric origin. A likely explanation for the complex structure of the core-level spectrum is the well-known, complex final-state effects in the core-level emission from this state, something that has been frequently encountered in connection with XPS from V oxides \cite{Gupta:1975aa,Sawatzky:1979,Panaccione:1997}.  Note that the detailed lineshape of the V~3p component might be more complicated than is described by the three peaks here, and that fitting with the broad and highly asymmetric grey peak is probably an oversimplification. However, none of these considerations is important for the data analysis here. Rather, the important conclusion is that the peaks are not due to geometrically different V emitters. The  V~3p spectrum for phase~I$'$ is very similar to that of phase I, both in peak positions and intensities. For phase II, on the other hand, the grey and purple components shift to significantly lower binding energy by $\approx$0.5~eV and the relative intensity of the  pink component significantly increases. 

\subsubsection{Structural Analysis by XPD} \label{XPDSubsection}

With the core-level spectral components fitted, XPD can be used to identify the different components in terms of their origins in geometrically different emitting atoms, and to determine the geometric structures of the three phases. The results are given in Fig. \ref{fig:XPD}, which shows XPD data for selected core-level components (bottom and top S~2p$_{3/2}$, and the low-binding-energy (pink) component of the V~3p peak). Stereographic projections of the modulation functions are shown with experimental data (orange) superimposed on the best-fit multiple-scattering simulations (greyscale). Sketches of the structural models are also given. After we had taken preliminary data to identify the origins of the different core-level components, we chose photon energies for acquiring the data sets in Fig. \ref{fig:XPD} such that the kinetic energy for emission from the top S would be below 150~eV.  The intent of using such a low energy was to enhance the cross section for backscattering from the underlying atoms, and thus to increase the sensitivity of the measurement towards the structural details of the layer. The photon energy for collecting data from the bottom S atoms, on the other hand, was chosen to be somewhat higher, in order to make use of the enhanced forward scattering at higher energies. 

We first discuss the structure of phase I.  As indicated above, LEED and STM results are compatible with either a 1H or 1T structure for this phase; however, since 2H bulk VS$_2$ has never been observed experimentally, one might speculate that phase~I is unlikely to be 1H. This is indeed confirmed by XPD: Fig. \ref{fig:XPD}(a)--(c) show simulations for the modulation functions of the 1H structure, while Fig. \ref{fig:XPD}(e)--(g) give the corresponding simulations for a structurally optimized 1T structure (layer thickness 2.9(1)~\AA~and lattice parameter 3.17(3)~\AA) along with a direct  comparison to the experimental data. The identification of the structure as 1T is confirmed by the excellent $R$-factors (shown in the bottom of each panel).  The poor agreement for the 1H phase is obvious, especially for the modulation function of the bottom S atom. It is easy to understand that the emission from this particular atom matters most for the distinction between the 1T and 1H phases: the relative coordination of the V and the top S atoms is the same for both structures, so neither the forward-scattering-dominated modulation function of V~3p nor the backscattering-dominated modulation function of the top S atom can be expected to be very different for the two structures. Note that the modulation functions for the 1H phase in Fig. \ref{fig:XPD}(a)--(c) were calculated using the same lattice parameter as for the 1T phase (in agreement also with LEED and STM data), but a different choice of lattice parameter in the simulation would not be expected to significantly impact the qualitative disagreement between the simulation and the experimental data. Interestingly, the good agreement shown here for the 1T phase is obtained from a model with a single rotational domain on the surface---\textit{i.e.,} a simulation with an absence of mirror-twin domains.  Such domains commonly \cite{Gronborg:2015,Lehtinen:2015aa} (but not always \cite{Bana:2018}) occur in the synthesis of SL TMDCs. That the sample consists primarily of a single domain-orientation is evident in the three-fold (rather than six-fold) symmetry of the diffraction patterns.  We note that recent theoretical studies have predicted that the 1H structure should be the stable structural phase of SL VS$_2$ in the temperature regime investigated here \cite{Kan:2015,Zhang:2013}, but we do not find any indication of this in the present study.  A possible reason could be the interaction with the substrate.

\begin{figure*}
   \includegraphics[width=1.0\textwidth]{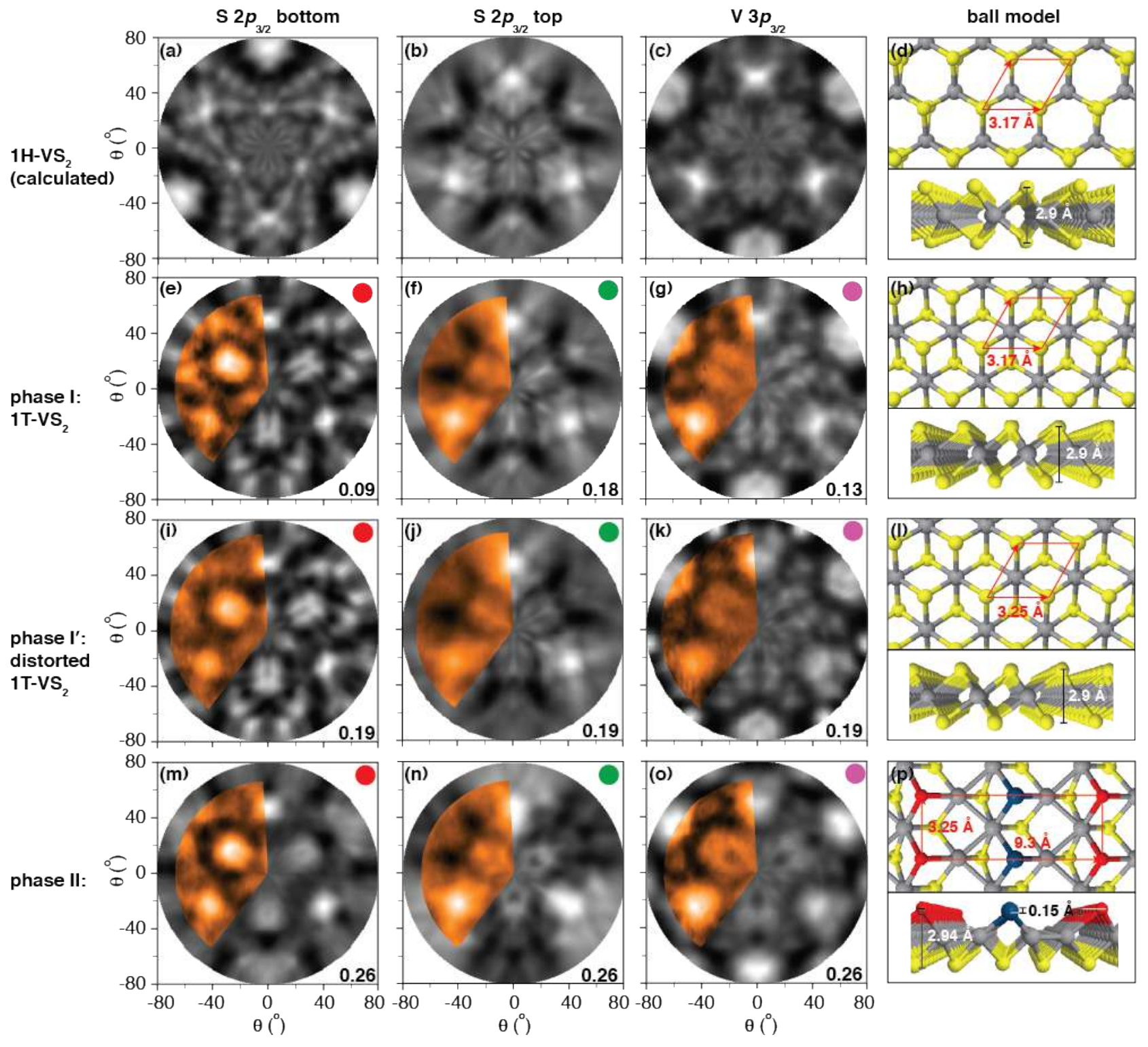}
   \caption{Stereographic projections of the modulation of the S~2p$_{3/2}$ bottom, S~2p$_{3/2}$ top, and V~3p$_{3/2}$ core level peaks, measured at $h\nu$~=~400~eV, 270~eV, and 270~eV, respectively.  Orange sectors are experimental data, while the grayscale patterns on which they are superimposed are corresponding XPD simulations.  The schematics at right---(d), (h), (l) and (p)---show the V$_{1+x}$S$_{2}$ structures used for the simulations.  (a)--(c) XPD simulations for 1H-VS$_{2}$ with a lattice constant of $a=3.17$~\AA{}.  (e)--(g) Experimental XPD patterns obtained from phase~I, superimposed on XPD simulations for 1T-VS$_{2}$ with a lattice constant of $a=3.17$~\AA{}.  (i)-(k) Experimental XPD patterns obtained from phase~I$'$, superimposed on XPD simulations for 1T-VS$_{2}$ with a lattice constant of $a=3.25$~\AA{}.  (m)--(o) Experimental XPD patterns obtained from phase~II, superimposed on XPD simulations for the lattice structure shown in (p).   In the case of phases I and I$'$, a single-domain orientation of the 1T configuration is used for the simulation, while for phase II the simulation includes three rotational domains probed simultaneously by the light spot. In (p) only a single rotational domain is illustrated.  In the schematics in (d), (h), (l), and (p), yellow spheres indicate S atoms and grey spheres, V atoms.  The blue spheres in (p) indicate down-modulated top-S atoms and the red spheres, up-modulated top-S atoms.  The coloured circles in the upper-right-hand corners of the data panels indicate which of the correspondingly-coloured XPS peaks in Fig. \ref{fig:XPS} is the source of each set of XPD data shown here; however, note that the data here was acquired at different photon energies than the XPS data in Fig. \ref{fig:XPS}. The number at the lower right corner of each panel is the $R$-factor associated with the simulation shown.}
   \label{fig:XPD}
\end{figure*}

The XPD data for phase~I$'$, Fig. \ref{fig:XPD}(i)--(k), sustain all the main features seen for phase I (Fig. \ref{fig:XPD}(e)--(g)) with only very minor changes.   The best agreement between XPD measurements and simulation is again found for the 1T  configuration (the simulation here also assumes single-domain orientation), still with a thickness of 2.90(1)~{\AA} and a slightly increased lattice constant of 3.25(5)~\AA. Note that a small increase of the in-plane lattice constant is consistent with the LEED results for the transition from phase I to phase I$'$.

Phase II represents an entirely different structure from the two other phases, as already indicated by LEED and STM. We have determined this structure  by simulating XPD modulation functions for a range of crystal structures, with the requirement that simulated structures must be consistent with STM and LEED data (\textit{i.e.,} must possess a rectangular $\approx$ 8.2~\AA{}$\times$3.1~\AA{} surface unit cell and exhibit an up-down modulation of the top S along the direction of the longer side of the unit cell), and minimizing the $R$-factor with respect to the experimentally obtained XPD data; furthermore, we required that simulated structures must be obtainable by a plausible phase transition involving S desorption from the known structure of phase I$'$ (Fig. \ref{fig:XPD}(l)).  Having a rectangular unit cell, phase II forms three domains on the three-fold Au(111) substrate, and these domains are rotated by 120$^{\circ}$ with respect to each other.  The presence of all three domains was taken into account in the simulations. The resulting optimized structure is shown in Fig. \ref{fig:XPD}(p).  Good agreement with the measurements was found for a simulated unit cell of size 9.30(12)~\AA{}$\times$3.25(7)~\AA{} and layer thickness 2.94(11)~\AA{}. The top of the layer consists of two rows of S atoms, as observed by STM, with a height difference of 0.15(13)~\AA{}, also consistent with STM. 

The overall agreement between simulated and measured modulation functions is very good for all three phases but is best for phase I. The higher overall $R$-factor for phase I$'$ compared to phase I is easily understood in terms of phase I$'$ being a S-deficient and hence defective structure. The even higher $R$-factor for phase~II might be related to the presence of domain boundaries between the rotational domains in phase~II (the boundaries are not accounted for in the simulations) and to local anti-phase domain boundaries such as the one seen in the inset of Fig. \ref{fig:STM-LEED}(e)---a frequently observed defect in the local atomic structure of this phase. 

A question so far not touched upon is the pathway for the formation of phase~II from phase~I. A possible scenario is proposed in Fig. \ref{fig:pIpIIreaction}. Starting from the intact phase~I in Fig. \ref{fig:pIpIIreaction}(a), a rectangular unit cell is shown that eventually transforms into the unit cell of phase~II. The transition to phase~II then proceeds by the desorption of the S atoms marked by black crosses in Fig. \ref{fig:pIpIIreaction}(b) and by a re-arrangement of the remaining atoms to form the structure of phase~II in Fig. \ref{fig:pIpIIreaction}(c). This rearrangement includes the contraction of the unit cell in the long direction and the distortion of the V lattice from a hexagonal geometry into zigzag chains.  These chains have V-V distances of only 2.45(4)~{\AA}, which is smaller than the nearest-neighbour distance in bulk metallic V in the body-centred cubic structure; this suggests that phase~II might exhibit metallic bonding upon sulphur depletion.   The transition pathway described here can easily explain several observations we have made in connection with phase~II.  A desorption of the S atoms from the top layer would, apart from being na\"{i}vely expected, explain the loss of photoemission intensity in the top-layer S core levels (even though such a loss is hard to quantify).

\begin{figure*}
\centering
   \includegraphics[width=0.9\textwidth]{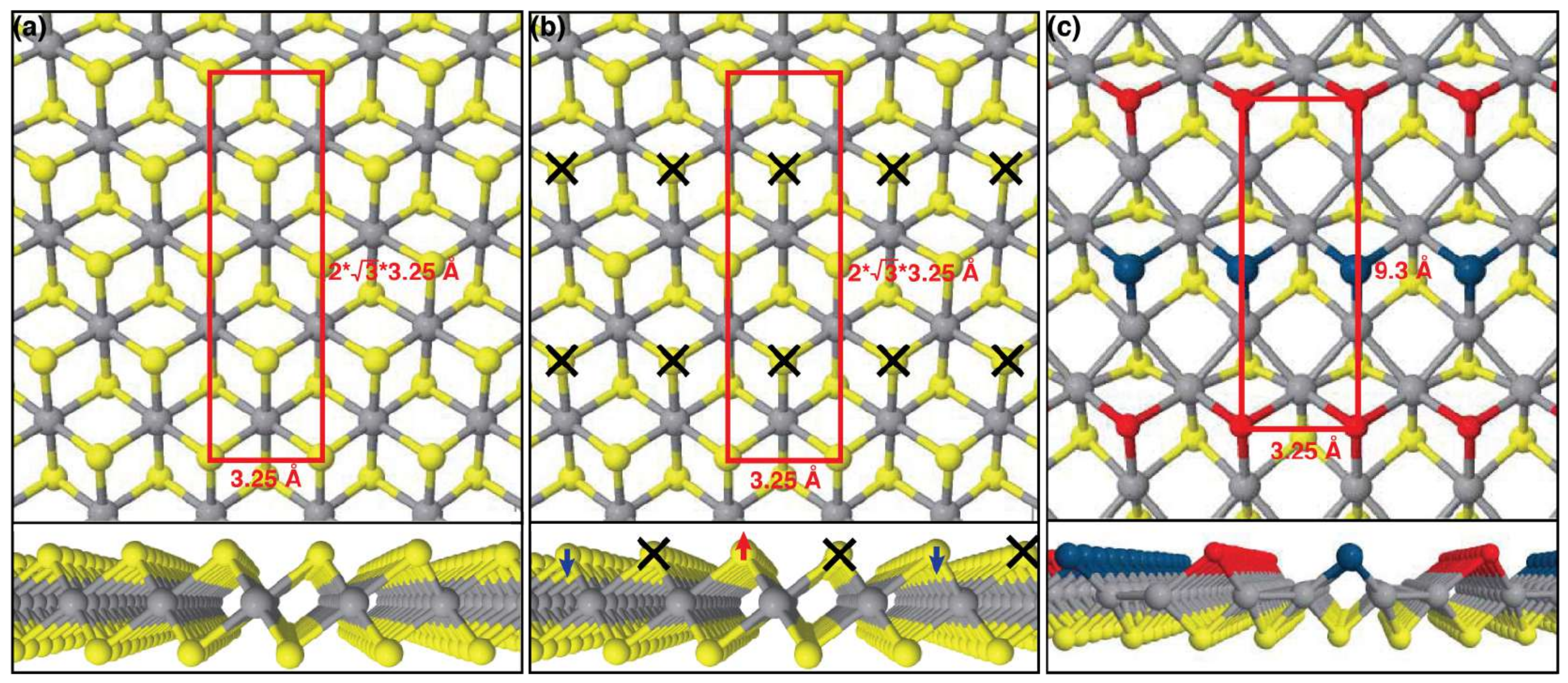}
   \caption{Possible reaction pathway from phase I to phase II. (a) Structure of phase I with a rectangular unit cell that forms the basis for the transformation to phase II. (b) The black crosses mark the S atoms in the unit cell that are lost in the transition. (c) Final structure of phase II that requires a distortion of the atoms in the layer and a shrinking of the unit cell in the long direction.  }
   \label{fig:pIpIIreaction}
\end{figure*}

\section{Conclusion}

In conclusion, we have reported here the first realization of two new single-layer V sulphide compounds.  One of these, 1T-VS$_2$, has a bulk analogue, even though the stoichiometric bulk VS$_2$ is metastable. It will be extremely interesting to probe the magnetic properties of the single layer compound and its susceptibility to charge density wave transitions, and to compare its behaviour to that of the recently synthesized and closely related material SL VSe$_2$ \cite{Bonilla:2018aa}. Heating the VS$_2$ layer in vacuum leads to a S-depleted structure with an increased in-plane lattice constant. This is consistent with the behaviour of bulk 1T-V$_{1+x}$S$_{2}$ \cite{Moutaabbid:2016}. 

The other new two-dimensional form of V$_{1+x}$S$_{2}$, called ``phase~II'' in this paper, is in many ways even more interesting, as it does not have a bulk analogue. The new material's electronic structure, possible charge density wave phases, and magnetic ordering remain to be explored, and an interesting open question is whether the S loss in the transition from stoichiometric SL VS$_2$ to the new phase modifies the material properties in such a way as to alter the magnetic properties of bulk VS$_2$ and its tendency to form charge density waves.

\section*{Acknowledgments}

The authors acknowledge helpful discussions with Albert Bruix and Alex Khajetoorians.  This work was supported by the Danish Council for Independent Research, Natural Sciences under the Sapere Aude program (Grants No. DFF-4002-00029 and DFF-6108-00409), by the VILLUM FONDEN via the Centre of Excellence for Dirac Materials (Grant No. 11744), and by the Aarhus University Research Foundation.
 



\end{document}